\documentclass[aps,amssymb,superscriptaddress,amsmath]{revtex4}

\usepackage{bm}
\usepackage{graphicx}

\begin{document}

\title{Dynamical symmetry breaking under core excitation in graphite:
Polarization correlation in soft X-ray recombination emission}

\author{Y. Harada}
\email{E-mail: harada@spring8.or.jp}
\author{T. Tokushima}
\author{Y. Takata}
\affiliation{RIKEN/Spring-8, Sayo-gun, Hyogo 679-5148, Japan}
\author{T. Takeuchi}
\affiliation{Tokyo University of Science, Shinjuku, Tokyo 162-8601, Japan}
\author{Y. Kitajima}
\affiliation{Photon Factory, Institute of Materials Structure 
Science, High Energy Accelerator Research Organization, Tsukuba, 305-0801, Japan}
\author{S. Tanaka}
\email{E-mail: stanaka@ms.cias.osakafu-u.ac.jp}
\affiliation{Department of Materials Science, Osaka Prefecture 
University, Sakai, 599-8531, Japan}
\author{Y. Kayanuma}
\affiliation{College of Engineering, Osaka Prefecture 
University, Sakai, 599-8531, Japan}
\author{S. Shin}
\affiliation{RIKEN/Spring-8, Sayo-gun, Hyogo 679-5148, Japan}
\affiliation{The Institute for Solid State Physics, University of Tokyo, Kashiwa, 277-8581, Japan}

\date{\today}

\begin{abstract}
\textbf{ABSTRACT}

A recombination emission spectrum is applied to study the 
local lattice distortion due to core excitation in graphite.
The recombination emission spectrum reveals a long low-energy
tail when the C $1s$ electron is excited to the $\sigma^*$ core 
exciton state. This indicates a large local lattice distortion around 
the excited carbon atom within a core hole lifetime ($\sim$10fs).
Theoretical calculation based upon an ionic cluster model well 
reproduces the experiments.
The strong polarization correlation between incident and emitted X-rays 
 is conclusive evidence of symmetry breaking in the core 
exciton state due to coupling with asymmetric vibrational modes.
\end{abstract}

\maketitle


Innovations in X-ray experiments have greatly 
enhanced our understandings of the relaxation 
dynamics of core excited states in many condensed systems.
Among them, spectroscopies of coherent second order processes,
such as resonant Auger electron emission and 
resonant X-ray emission spectroscopies, once believed to be difficult to 
obtain have now become common and the best tool to study 
ultrafast relaxation dynamics of the core excited 
states~\cite{Gelmukhanov99,Rubbenson00,Kotani01,Bruhwiler02}.

Since the core excited atom can be regarded as an electronically 
equivalent atom with the valence increased by 1 ($Z+1$ approximation), the 
sudden change of the charge balance around the excited atom 
may bring about many (shake-up or shake-off) satellites in the core 
level spectra.  
While electronic relaxation on core excitation has been much 
discussed, not much attention has been paid to the 
relaxation dynamics through electron-phonon coupling in the core 
excited states, especially in solid states.
This is because many have long believed that the time scale of the 
phonon relaxation, which usually falls in sub picoseconds, is far 
longer than a core hole lifetime (less than several femtoseconds), so 
that core 
excited state immediately decays with electronic Auger decay process
before heavy atoms start to move.
In simple molecules and adsorbates on solid surface, however, recent 
experiments and theoretical analyses of resonant Auger and soft X-ray 
emission spectroscopies have clearly indicated a large atomic 
displacements in core excited states when  a core electron is excited 
to a bound unoccupied state~\cite{Rubbenson00,Ueda00}.  

Meanwhile in solid states, there have been no reports on a large 
atomic displacement in a core excited state since the pioneering experiment 
of soft X-ray recombination emission in 
diamond and graphite 
done by Y. Ma {\it et al.}~\cite{Ma93}.
In the experiments, the soft X-ray recombination emission spectrum
showed a long low-energy tail when the C $1s$
core electron is excited into a bound 
core excited state, indicating a large lattice distortion  around
an excited 
atom~\cite{Ma93} in the core exciton state.
According to $Z+1$ approximation C $1s$ core exciton can be regarded 
as a nitrogen impurity atom well known to form a deep level in the 
middle of band gap 
with a large off-center instability
~\cite{Jakson90, Mainwood94, Mauri95}.  
Tanaka and Kayanuma interpreted the characteristic feature of 
the recombination emission spectra of diamond with a theory 
in which local phonon 
modes are strongly coupled with the quasi-degenerate core exciton
states~\cite{Tanaka96}.
They clarified that the off-center instability is 
induced by a cooperation of quasi-Jahn-Teller and Jahn-Teller effects 
in the core excited state, and predicted that the spectrum should 
have a distinct polarization correlation between the incident and 
emitted X-ray 
photons. This polarization correlation is decisive evidence that
the local symmetry is 
broken by ionic coupling of a core exciton with lattice 
phonon systems, i.e. {\it dynamical symmetry 
breaking } in core excited states.
Our aim here is to experimentally reveal the lattice relaxation dynamics in the core 
excited state of a solid state with soft X-ray recombination emission 
spectrum.


In this Letter, we report the recombination emission in graphite 
which clearly shows low-energy tail in the emission spectrum.
Considering the fact that the electronic 
configuration of the final state is just the same as in the initial 
state 
in the recombination emission process, the only possible mechanism 
responsible for 
this low-energy tail is the 
phonon relaxation in the intermediate core excited 
state.
Thus this should be the direct evidence of the large distortion 
around the excited atom in 
the core exciton state in graphite. 
This is quite in contrast to the case of resonant Auger emission 
process,  
where the final state interaction as well plays an important role to 
determine the spectral feature~\cite{Bruhwiler02}, 
which makes the interpretation of the spectrum complicated 
and sometimes leaves ambiguity.
In addition, we report here for the first time a strong polarization 
correlation of the recombination emission, which reveals the dynamical 
symmetry breaking by the ionic couplings in the core excited 
state.

Graphite has regained great attention since the discovery of carbon 
nanotubes that show wide unique physical properties with slight 
change of wrapping a single graphite layer. 
So the study on the effect of the electron-phonon interaction in graphite 
is important to understand the physical properties of nanotubes.


C $1s$ absorption and emission of highly oriented pyrolytic graphite 
(HOPG) 
was performed using a flat field soft X-ray emission 
spectrometer~\cite{Tokushima02} 
newly constructed in BL27SU at SPring-8~\cite{Ohashi01}. 
The energy resolution of incident and emitted photons are  less than 
0.1eV and 0.6eV at C $1s$ edge, respectively. 
The excitation energy was carefully calibrated using the energy 
position of Au $4f$ photoemission lines. 

\begin{figure}
\includegraphics[bb=0 150 500 700, width=8cm,clip]{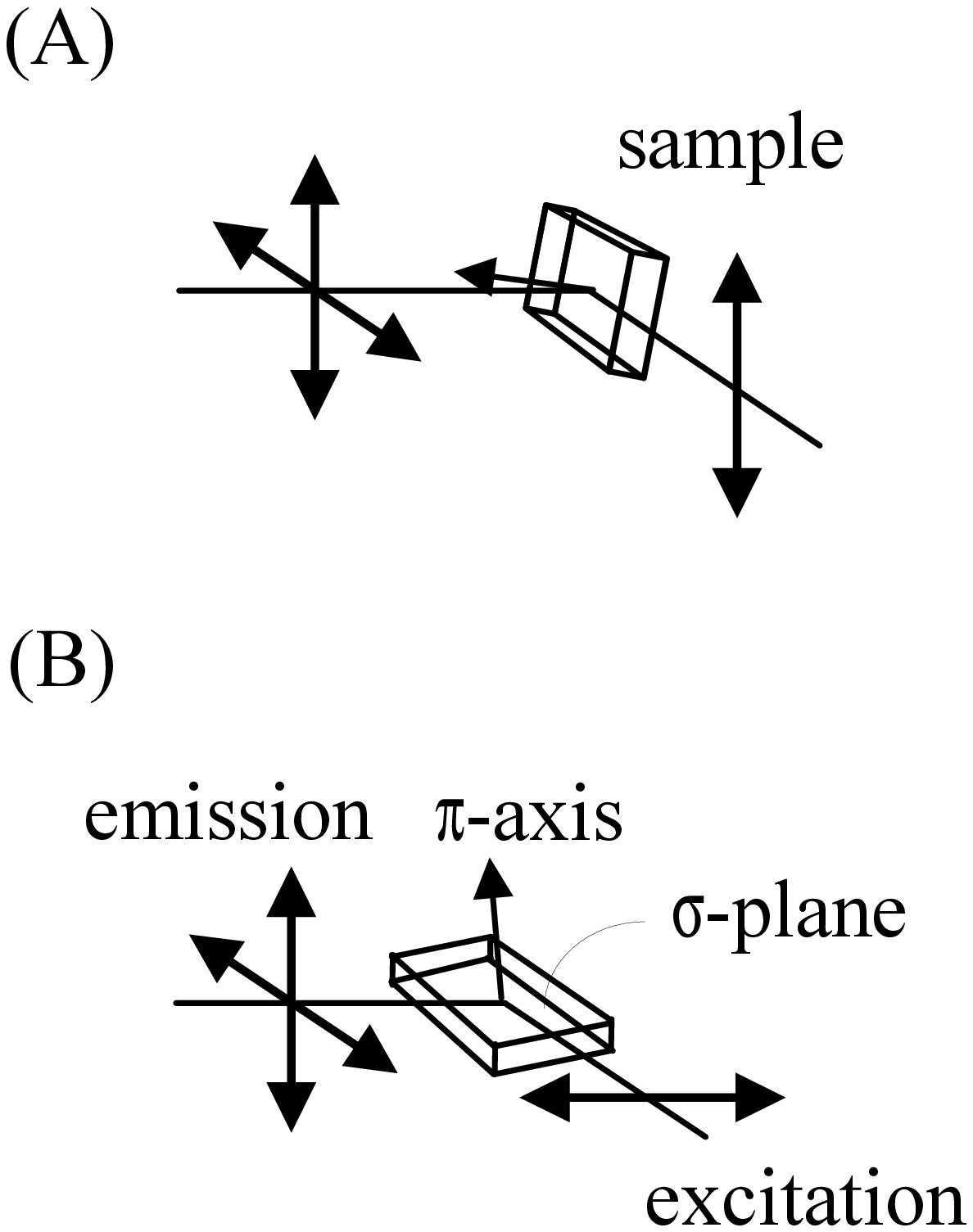}
    \caption{Experimental geometry; (A) polarized configuration; 
    (B) depolarized configuration. Two-headed arrow indicates the
    polarization vector of incident and emitted photons.}
\label{Fig1}
\end{figure}%

\begin{figure}
\includegraphics[bb=0 300 400 850, width=10cm]{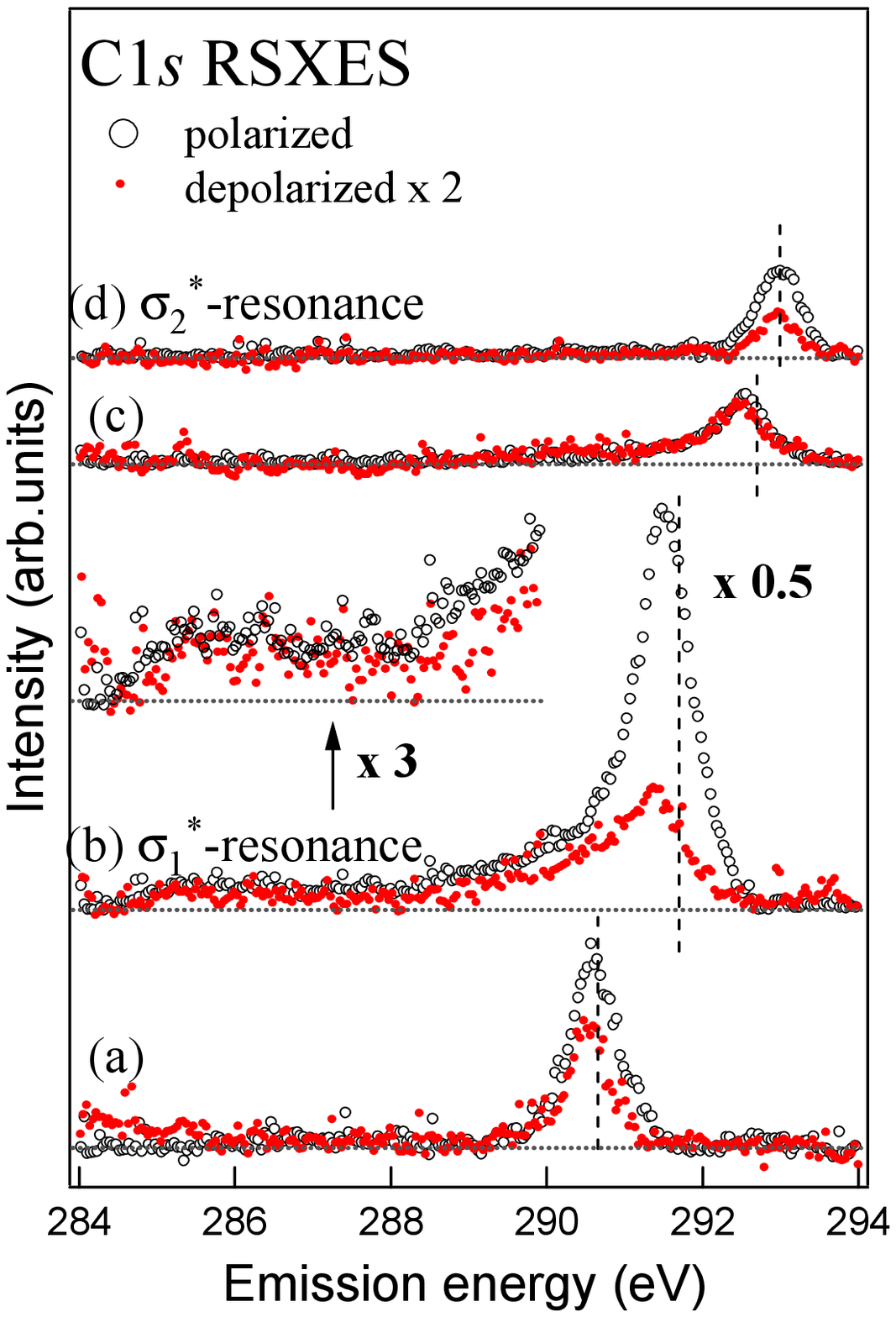}
    \caption{C $1s$ resonant soft X-ray emission spectra of graphite.
    The excitation energies are shown in the absorption spectrum in the 
    inset; they are tuned at $\hbar\Omega_{1}=$ (a) 
290.7eV, (b) 291.7eV, (c) 292.7eV, and (d) 293eV.    Dashed vertical lines 
indicate the position of the excitation 
energy.    Polarized and depolarized spectra are drawn with the line with 
open circle and thick red line, respectively.
    }
    \label{Fig2}
\end{figure}

C $1s$ recombination emission spectra were measured at (A)'polarized' and 
(B)'depolarized' configurations as shown in Fig.~\ref{Fig1}.
While two independent polarizations of an emitted X-ray in the polarized
configuration are  
parallel and orthogonal to an incident X-ray polarization, they
are necessarily orthogonal to that of the incident X-ray in the depolarized 
configuration.
We show  the C $1s$ recombination emission obtained by 
scanning the excitation X-ray energy across $\sigma^*$ 
absorption band in Fig.~\ref{Fig2}. The valence x-ray 
emission band appears below 284eV~\cite{C1sXES}, which is not shown here.

The C $1s$ X-ray absorption spectrum (C $1s$ XAS) shown in the inset is attributed to the 
transition of a C $1s$ core electron to the antibonding $\sigma^*$ conduction states.
The peak at 291.7eV has been interpreted as the transition to a 
localized core exciton state ($\sigma_{1}^*$), which is followed by 
the broad absorption band attributed to the excitation to the 
delocalized scattering state ($\sigma_{2}^*$)~\cite{Batson93, Bruhwiler95, McCulloch96}.
The excitation energies $\hbar\Omega_{1}$ are indicated in the absorption spectrum: 
(a) 290.7eV, (b) 
291.7eV, (c) 292.7eV, and (d) 
293eV.
In Fig.~\ref{Fig2}, the emission spectra 
 in the polarized and depolarized configurations are shown 
by a line with open circle and a thick red line, respectively. 

The recombination emission spectra strongly depend on the excitation energy.
The spectrum has a long tail starting from the 
elastic line (Rayleigh line) towards the low energy side when $\hbar\Omega_{1}$ is
tuned with $\sigma_{1}^*$ core exciton peak 
(Fig.~\ref{Fig2}(b)).
This low energy tail almost disappears when $\hbar\Omega_{1}$ is 
1.0eV off-resonant below the $\sigma_{1}^*$ resonance 
(Fig.~\ref{Fig2}(a)).
Similarly when $\hbar\Omega_{1}$ is increased, the low energy tail is 
shortened (Fig.~\ref{Fig2}(c)), and at the $\sigma_{2}^*$ excitation, 
we cannot see any asymmetry in the 
spectral shape (Fig.~\ref{Fig2}(d)). 
The appearance of this low energy tail is interpreted as a {\it 
hot luminescence}, the photon emission taking place as the phonon 
relaxation proceeds~\cite{Tanaka96}: The emitted soft X-ray photon energy is
gradually lowered as a part of electronic excitation energy  
being transferred to the vibrational subsystems.
When the excitation becomes off-resonant 
(Fig.~\ref{Fig2}(a)), 
the whole recombination emission process from the initial to final states becomes 
coherent and the intermediate core exciton state is virtually passed 
through.
The effective time during which the core excited state can be 
effectively coupled with phonon systems becomes shortened, resulting in the 
disappearance of the low-energy tail~\cite{Gelmukhanov99}.
In addition, we have found out a small hump around 285eV in Fig.~\ref{Fig2}(b).
This is attributed to the recombination emission when an induced coherent vibrational 
wave packet turns back at a turning point on an adiabatic potential surface of the core 
exciton state~\cite{Mahan77,Kayanuma88}.
These structures are 
hardly seen in the resonant Raman scattering spectrum in optical 
region, because the lifetime of the excited state is usually so long 
that the ordinary luminescence from the relaxed excited state dominates 
over these faint structure. 
Since it is these hot luminescence spectral features that well reflect 
the very early stage of the  
relaxation process~\cite{Mahan77,Kayanuma88}, 
the soft X-ray recombination emission is powerful tool to study the 
ultrafast relaxation dynamics of the 
core excited state.

The most intriguing features are the polarization correlation  in Fig.~\ref{Fig2}.
The following two points should be stressed here; (i) The low-energy tail 
in recombination emission has been observed both in depolarized as well as  
polarized configurations; (ii) The intensity of the low-energy tail relative
to the integrated intensity of the recombination emission on the $\sigma_{1}^*$ core excitation is more pronounced in depolarized configuration than in
polarized configuration.
These suggest that the local symmetry around the excited atom 
is broken under the $\sigma_{1}^*$ core excitation.
Considering a minimal cluster consisting of four C atoms ----- at the 
center is the excited atom surrounded by three nearest neighbor atoms
----- we have three stretching vibrational modes as major coupling 
vibrations with the core exciton state: one symmetric and two 
asymmetric stretching vibrational modes.
Since the depolarization process requires at least one symmetry 
breaking phonon emission, the asymmetric vibrations must be induced 
under $\sigma^*$ 
core excitation through the ionic coupling, while the coupling with 
the symmetric vibration does not change the structural symmetry.
A theoretical analysis of these spectral shapes 
enables us to evaluate the coupling strengths of the core exciton with 
these vibrational modes.

Based on the cluster model, we consider a three 
antibonding $sp^2$ hybridized orbitals  
as an electronic basis set which are coupled with the three stretching 
vibrations~\cite{Ueda00}.
In a representation of a symmetrized basis set, the Hamiltonian for
the core exciton 
states is represented by 
\begin{eqnarray}
    \label{Hamiltonian}
    H_{e}=\left[
    \begin{array}{ccc}
	\epsilon_{s}-(\alpha/\sqrt{3}) Q_{s} 
	& -(\beta/\sqrt{3}) Q_{x} 
	& -(\beta/\sqrt{3}) Q_{y} \\
	-(\beta/\sqrt{3}) Q_{x}
	& \epsilon_{p}-(\alpha/\sqrt{3}) Q_{s} - (\beta/\sqrt{6}) Q_{y} 
	& -(\beta/\sqrt{6}) Q_{x}  \\
	-(\beta/\sqrt{3}) Q_{y} 
	&  -(\beta/\sqrt{6}) Q_{x} 
	& \epsilon_{p}-(\alpha/\sqrt{3}) Q_{s} + (\beta/\sqrt{6}) Q_{y} \\
    \end{array}
    \right] +H_{0}   ,
\end{eqnarray}
with 
\begin{equation}
    H_{0}=\hbar \omega_{s}b_{s}^\dagger b_{s}+\sum_{i=x,y}\hbar 
    \omega_{p}b_{i}^\dagger b_{i}   \;,
\end{equation}
where $H_{0}$ is unperturbed Hamiltonian for the vibrational systems; 
$b_{i}$ ($b_{i}^\dagger$) is boson annihilation (creation) 
operator for each mode: $Q_{i}=1/\sqrt{2}(b_i+b_i^\dagger ) \; (i=s,x,y)$;
$\hbar \omega_{s}$($\hbar \omega_{p}$) is the energy of the vibrational system in $s$ ($p=x,y$) mode. The energies of the $s$ and $p$ states at ${\bf Q}={\bf 0}$ are denoted by $\epsilon_s$ and $\epsilon_p$, respectively, and $\alpha$ 
($\beta$) is the coupling strength with the symmetric  
(asymmetric) modes.
Because of the lack of inversion symmetry, the coupling with asymmetric 
modes $Q_{x}$ and $Q_{y}$ brings about simultaneously the Jahn-Teller coupling 
within $p$ symmetry core exciton states and the quasi-Jahn-Teller 
coupling between the $s$- and $p$-symmetry core exciton states.

\begin{figure}[b]
\includegraphics[bb=100 300 450 650, width=10cm, clip]{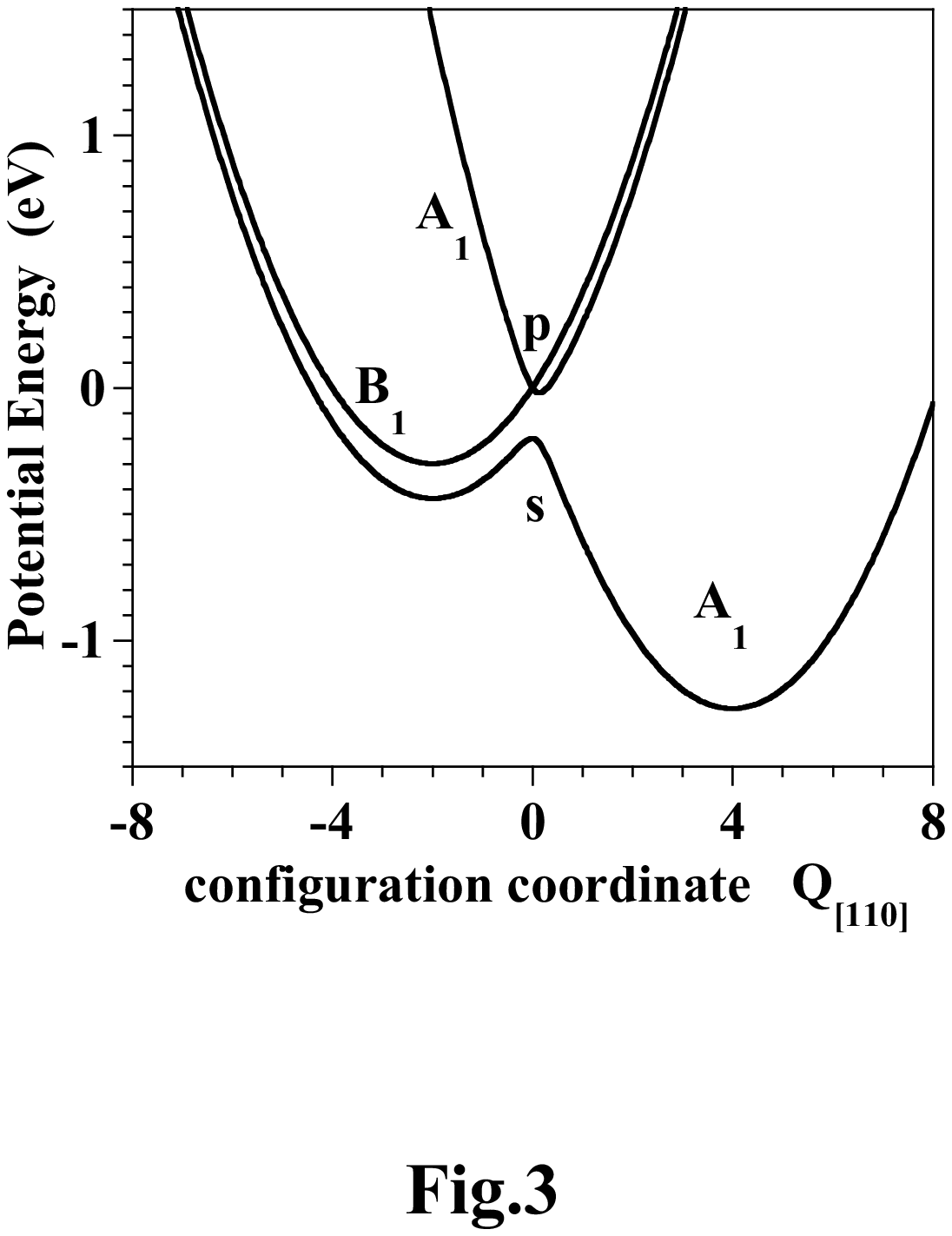}
    \caption{The calculated adiabatic potential surface in the core 
    excited state along a bond direction.}
    \label{Fig3}
\end{figure}%

\begin{figure}[t]
\includegraphics[bb=100 300 450 680, width=10cm, clip]{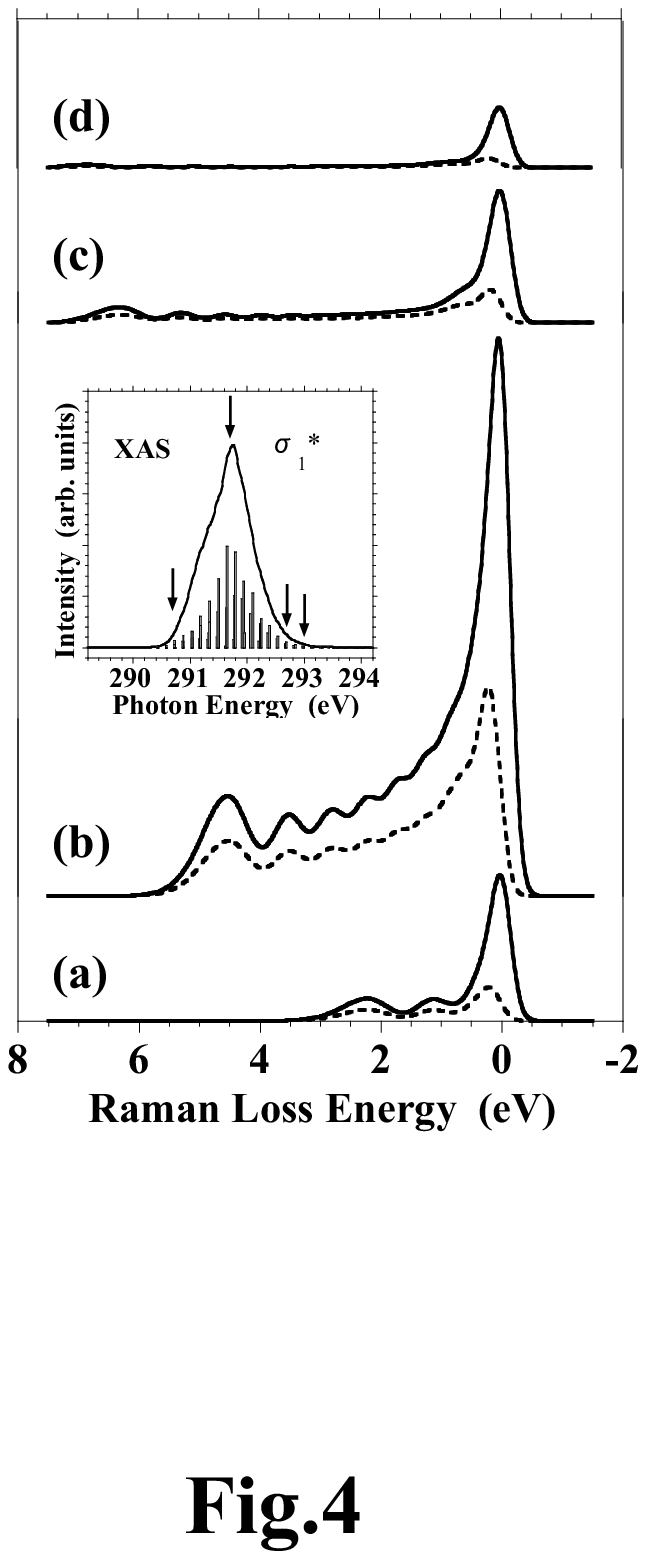}
    \caption{The calculated results of resonant soft X-ray emission 
spectra across the $\sigma^*$ absorption peak. 
Polarized and depolarized spectra are drawn with the line and dashed line, 
respectively. The intensity of the depolarized spectra is multiplied by 2.
The X-ray absorption spectrum is shown in the inset.
}
    \label{Fig4}
\end{figure}%

In Fig.~\ref{Fig3}, we depicted an aspect of adiabatic potential 
surfaces in the core excited state for a set of parameters shown below.
The horizontal axis represents the atomic displacement of the  
excited atom along a bond direction.
The unit of the horizontal axis is then evaluated to be $\sqrt{
(\hbar/M\omega_{s,p})}$, where $M$ denotes the mass of carbon atom.
Doubly degenerated $p$ symmetry core exciton states at $Q=0$ is split 
into $A_1$ and $B_1$ states by 
Jahn-Teller coupling, and the lowest branch for the $s$ symmetry core 
exciton state is greatly lowered by the quasi-Jahn-Teller coupling with 
the upper branch.
For symmetry reasons, one of the upper 
branches can couple with the lowest branch.
It is clearly seen that as a result of the quasi-Jahn-Teller coupling the 
on-center position becomes unstable and the 
equilibrium position is shifted about 0.2{\AA} away from the origin with about 
1.27eV relaxation energy.
The C $1s$ core electron is excited to the dipole allowed $p$ symmetry core 
exciton states, at $Q=0$ of the upper branch, and a nonadiabatic coupling 
causes it decay 
into the 
lowest branch.
The low energy tail in the recombination emission is attributed to the
relaxation process going along the lowest 
branch toward the equilibrium, and the hump structure is due to the
hot luminescence
at the turning point. 
As the lattice relaxation proceeds along with the potential
surface of the asymmetric modes, 
the local symmetry around the excited atom is broken down, 
which causes the observed prominent low-energy tail structure in the depolarized 
configuration.  

We show the calculations of the recombination emission in Fig.~\ref{Fig4}, 
which is calculated fully quantum mechanically by the formula for a 
second order optical process~\cite{Tanaka96}.
We have determined the parameter values so that the experiments 
can be well reproduced:
$\hbar \omega_{s}=\hbar \omega_{p}=0.15$eV, 
$\epsilon_{p}-\epsilon_{s}=0.4$eV, $\alpha=0$, $\beta=4.9$ in unit 
of $\hbar\omega_{p}$, 
and $\Gamma_{1s}=45$meV.
In our model, the damping of phonon oscillation is neglected for
simplicity. This approximation is justified because the core excited state
decays before reaching the thermal equilibrium, and the x-ray emission
takes place only within that short lifetime of the core hole.
The recombination emissions are plotted for the same excitation energies in the
experiments.
The calculated absorption spectrum is also shown in the inset, and the 
excitation energies are indicated by arrows.
The calculations satisfactorily explain the experiments.
The low energy tail is prominent when the excitation energy is tuned with the 
$\sigma_{1}^*$ peak both in the polarized and the 
depolarized configurations.
The relative intensity of the low-energy tail is more pronounced in the 
depolarized configuration than in the polarized configuration because the 
strong Rayleigh component is forbidden in 
the depolarized configuration.
The couplings with the asymmetric modes are essential to 
reproduce the experiments; In fact the coupling with 
the symmetric mode has been neglected here.
This is quite in contrast to the case of a BCl$_{3}$ molecule with a same 
point symmetry, $D_{3h}$, where the coupling with 
symmetric stretching mode plays an important role~\cite{Ueda00}.
This may reflect a qualitative difference of the vibrational coupling effects
between molecules and crystals; 
In graphite the expansion of the cluster may be greatly suppressed by the 
existence of surrounding atoms.

Under the higher excitation (Fig.~\ref{Fig4}(c) and (d)), asymmetry in 
the recombination emission disappears consistently with the experiments, 
for the 
reason mentioned above~\cite{Gelmukhanov99}.  In reality, 
when the $\hbar\Omega_{1}$ is well into high energy 
continuum, we should take into account itinerancy of the
excited electron over the entire crystal. The delocalization of the 
excited electron makes the exciton-phonon 
coupling suppressed~\cite{ToyozawaBook}, letting the emission peak symmetric.

In conclusion, our experiments of the recombination emission confirm a large local lattice
distortion around an excited atom in the core exciton state in graphite.
A remarkable polarization correlation between 
incident and emitted X-rays has been observed for the first time, leading to 
the conclusion that the dynamical symmetry breaking in the core exciton 
state is mainly due to the coupling of the core exciton states with 
asymmetric vibrational modes. 

We acknowledge for valuable technical assistance from the staff of BL27SU in SPring-8. We also thank Dr. Chainani for his critical reading of the manuscript and helpful comments.
This experiment was carried out with the approval of the SPring-8 Proposal 
Review Committee (2002B0675).


\end{document}